\documentclass{IEEEtran}
\usepackage{cite}
\usepackage{amsmath,amssymb,amsfonts}
\usepackage{graphicx}
\usepackage{textcomp,nicefrac}
\usepackage{subfigure}
\usepackage{multirow}

\def\BibTeX{{\rm B\kern-.05em{\sc i\kern-.025em b}\kern-.08em
T\kern-.1667em\lower.7ex\hbox{E}\kern-.125emX}}
\begin{document}
\title{Machine-Learning-Based Waveform Discrimination in the Front-End Electronics of the Belle II Central Drift Chamber for Cross-Talk Noise Reduction}
\author{Yun-Tsung Lai, \IEEEmembership{Member, IEEE}, Taichiro Koga, Yu Nakazawa, Nanae Taniguchi, Keisuke Yoshihara
\thanks{Y. T. Lai, T. Koga, Y. Nakazawa, and N. Taniguchi are with High Energy Accelerator Research Organization (KEK), Ibaraki 305-0801, Japan
 (e-mail: ytlai@ post.kek.jp).
 K. Yoshihara is with University of Hawaii, Honolulu, HI 96822, USA.}
}

\maketitle

\begin{abstract}
Machine learning (ML) inference on FPGAs has been widely adopted in real-time triggering of collider experiments for detector signature identification. In contrast, the ML application in Front-End Electronics (FEE) has not yet been fully explored, primarily due to constraints such as limited FPGA resources, power consumption, and localized detector coverage of individual FEEs.

In this work, we develop an ML-based waveform discrimination method for the Central Drift Chamber (CDC) of the Belle II experiment to suppress cross-talk noise at the front-end level. The Belle II CDC is a key charged-particle tracking detector for both offline and the real-time hardware trigger. During Belle II operation, background wire hits have been observed in the CDC FEE, where multiple hits occur in neighboring anode wires by large energy deposit. The hardware track trigger employs a Hough transformation based on track segments formed by combining hits from multiple wire layers. Due to the reduced information, the hardware track trigger is sensitive to cross-talk noise, hence resulting in an increased fake trigger rate with higher luminosity in the future.

We employ compact and fast Boosted Decision Tree (BDT) models implemented in a Xilinx Virtex-5 FPGA of the CDC FEE, where waveform classification is performed independently for each wire channel in a fully pipelined architecture. Offline studies using recorded waveform data show that the cross-talk noise can be reduced by approximately a factor of two while maintaining a signal efficiency above 98\%. The deployed firmware was further validated during dedicated Belle II calibration runs, demonstrating reductions of up to 50\% in track segment and track trigger rates while preserving the trigger acceptance for events containing reconstructed charged tracks within 10

This work demonstrates the technical feasibility of compact and low-latency ML inference in detector FEE and highlights its potential for future intelligent detector readout systems in high-energy physics experiments.

\end{abstract}


\section{Introduction}
\label{sec:introduction}
\IEEEPARstart{A}{s} the successor to the Belle experiment~\cite{belle} and the KEKB collider~\cite{kekb1,kekb2}, the upgraded Belle II experiment~\cite{belle2} and the SuperKEKB collider~\cite{superkekb} were designed to explore physics beyond the Standard Model through precision measurements and searches for rare processes with a larger amount of data. With the employment of the nano-beam scheme~\cite{nano}, SuperKEKB aims for a peak instantaneous luminosity of $6\times 10^{35} \textrm{cm}^{-2}\textrm{s}^{-1}$ and an integrated luminosity of 50 ab$^{-1}$. More precise measurements and studies on the rare $B$ meson decays from the $\Upsilon(4S)\to B\overline{B}$ events in $e^{+}e^{-}$ collision is one of the primary physics goals of Belle II. In addition, Belle II conducts a broad physics program including charm meson and $\tau$ lepton physics, dark sector searches, and physics beyond the Standard Model. Up to May 2026, Belle II has recorded an integrated luminosity of about 820 fb$^{-1}$.

The Belle II spectrometer is designed as a general-purpose detector for comprehensive and precise particle detection and identification. It consists of seven sub-detectors. The Pixel Detector (PXD), Silicon Vertex Detector (SVD), and Central Drift Chamber (CDC) provide charged particle tracking, vertex reconstruction, and momentum measurement. The Time-Of-Propagation (TOP) counter in the barrel region and the Aerogel Ring-Imaging Cherenkov (ARICH) at the forward end-cap region provide charged-hadron identification, particularly for kaon--pion separation. The Electromagnetic Calorimeter (ECL), composed of CsI(Tl) crystals, measures electromagnetic showers from photons and electrons. A magnetic field of 1.5~T is generated by a superconducting solenoid surrounding the ECL. The outermost detector, the KL and Muon Detector (KLM), consists of resistive-plate chambers and plastic scintillators for muon identification and $K_{L}$ detection.

In addition to the detector subsystems, Belle II includes a Level-1 (L1) trigger system~\cite{trg1} and a Data Acquisition (DAQ) system~\cite{daq1,daq2} for real-time event processing, readout, and storage. The Front-End Electronics (FEE) devices of each detector digitize analog signals from the detectors and transmit the resulting data to the L1 trigger and DAQ systems. To reduce the amount of data recorded by the DAQ system, the L1 trigger system performs real-time event reconstruction and selection using information from the FEEs of CDC, ECL, TOP, and KLM. The L1 trigger system is implemented using approximately 100 Field-Programmable Gate Arrays (FPGAs) and the trigger decision is required to be made within a latency of 4.4 $\mu$s. The selected events are subsequently collected and assembled by the DAQ system.

CDC~\cite{cdc}, which is one of the tracking devices in Belle II, provides input data for both offline reconstruction and real-time L1 trigger tracking. Since the start of Belle II operation in 2018, the CDC FEEs~\cite{cdcfee} have been affected by a cross-talk noise phenomenon that generates spurious hit signals in the anode wires and affects trigger tracking performance. Compared with the offline tracking, the trigger tracking uses reduced input information, making it particularly sensitive to such noise-induced hits, which can lead to fake track candidates and increased trigger rates.

In this work, we investigate the use of machine-learning (ML) methods to discriminate the cross-talk noise from signals induced by charged particles using waveform information available in the CDC FEEs. Unlike the trigger system, which receives only digitized hit and timing information, the FEE has access to the full ADC waveform, which provides unique information for noise suppression. The ML-based waveform classifier is implemented in the FPGA chip of the CDC FEEs to identify and suppress the noise before transmission to the trigger system. To the best of our knowledge, this is one of the pioneering demonstrations of ML inference in detector FEE for real-time processing in experimental particle physics.
This approach demonstrates the feasibility of deploying compact and low-latency ML algorithms in detector FEE devices and provides a potential framework for future upgrades of ML-based intelligent detector systems. 

We organize this paper as follows. Section~\ref{sec:cdc} introduces the CDC detector, its FEE device, and the cross-talk noise phenomenon.
Section~\ref{sec:ml} describes the development of the ML models based on collected CDC waveform data and their implementation in CDC FEE's FPGA chip.
Section~\ref{sec:perf} summarizes the performance validation based on both waveform data and experimental operation.
Finally, Section~\ref{sec:prospect} discusses the prospects of ML applications in detector FEEs. Section~\ref{sec:conclusion} concludes this work.

\section{Belle II Central Drift Chamber}
\label{sec:cdc}

\subsection{Structure of Belle II Central Drift Chamber}
\begin{figure}[t]
\centerline{\includegraphics[width=3.0in]{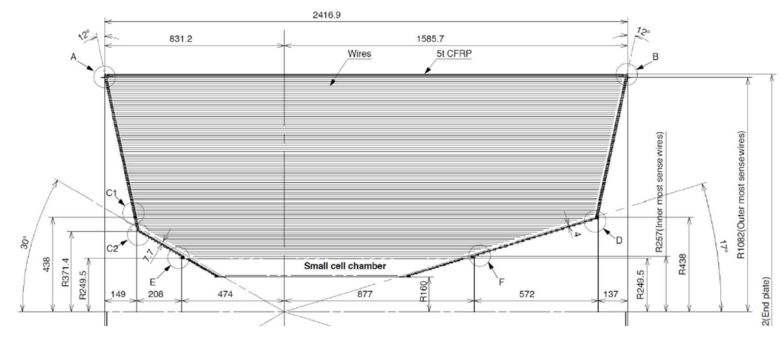}}
\caption{The mechanical design of the Belle II CDC detector.}
\label{fg:cdc}
\end{figure}

\begin{figure}[t]
\centerline{\includegraphics[width=2.0in]{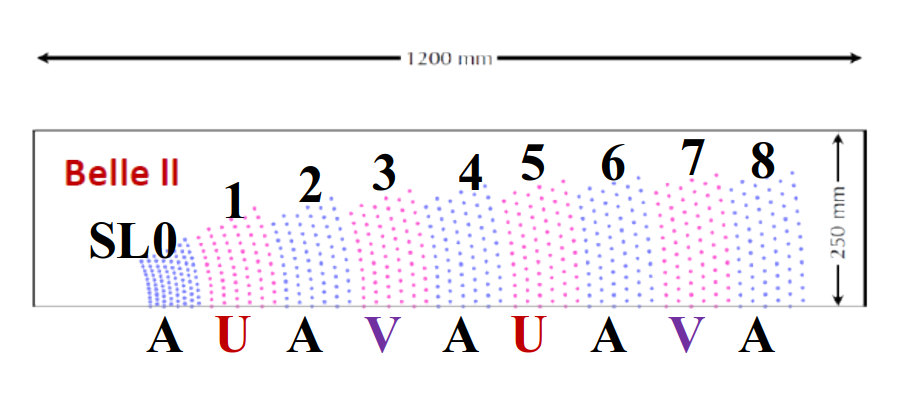}}
\caption{Wire configuration of the Belle II CDC detector.}
\label{fg:cdcwire}
\end{figure}

The Central Drift Chamber (CDC)~\cite{cdc} is one of the major charged tracking devices of the Belle II experiment. Figure~\ref{fg:cdc} shows a longitudinal view of the CDC's structure. It contains 14,336 anode wires operated with a mixture of He and ethane as the ionization gas. All the anode wires are organized into 56 wire layers and grouped into nine super-layers (SLs). The innermost eight wire layers form a standalone small-cell chamber~\cite{smallcell}, denoted as SL0, and each of the remaining SLs (SL1--SL8) consists of six wire layers. The anode wires are arranged in a staggered geometry, in which neighboring wire layers are shifted by half a cell width. The radial cell size is 0.8 cm in SL0 and 1.0 cm in the outer SLs. This configuration improves the left--right ambiguity resolution. Figure~\ref{fg:cdcwire} illustrates the wire configuration of the CDC of each SL, where the axial SLs (A) contain anode wires parallel to $z$-axis (defined as the SuperKEKB electron beam's direction) and the stereo SLs (U and V) are tilted by small stereo angles in opposite directions. Based on simulation studies, such an alternating wire configuration among the nine SLs enables three-dimensional track reconstruction in the L1 trigger system with a more precise measurement on the longitudinal track position~\cite{belle2}.

\subsection{Front-End Electronics of CDC}
\begin{figure}[t]
\centerline{\includegraphics[width=2.0in]{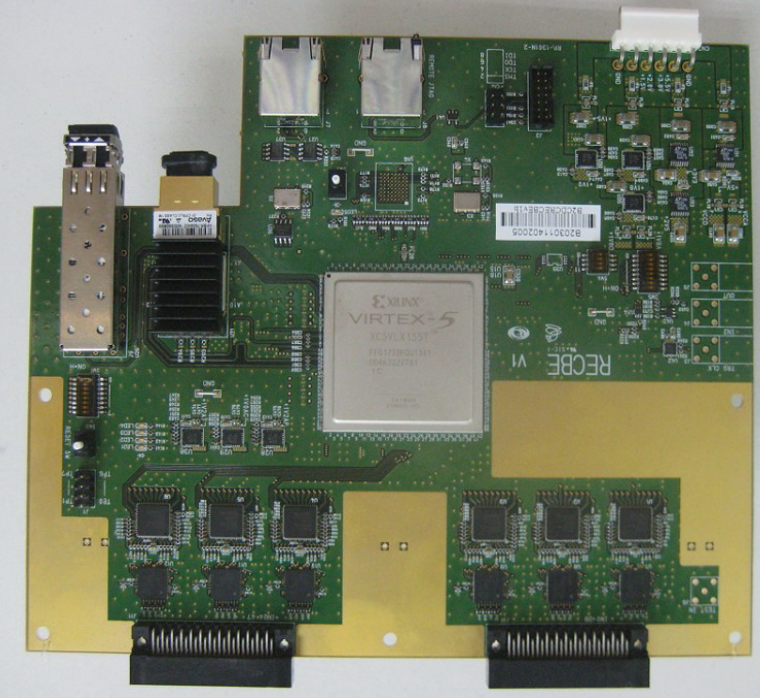}}
\caption{Front-End Electronics device of Belle II CDC.}
\label{fg:fee}
\end{figure}

The Front-End Electronics (FEE) device~\cite{cdcfee} of the CDC is shown in Fig.~\ref{fg:fee}. Each FEE board processes 48 anode wire channels. The analog signals from these wires are processed by six custom Amplifier-Shaper-Discriminator (ASD) ASICs and six Analog-to-Digital Converter (ADC) ASICs, where each ASIC handles eight wire channels.
The ASD ASIC contains a current-sensitive preamplifier, shaping and baseline-restoration circuits, an analog buffer for waveform sampling, and a comparator for timing measurement. The analog waveform is digitized by ADCs at 31.4~MSPS for waveform sampling and charge measurement, while the comparator output is digitized by a Time-to-Digital Converter (TDC) in a Xilinx Virtex-5 XC5VLX155t FPGA~\cite{virtex5}. Each wire hit is represented by a one-bit hit flag and a TDC timing value encoded as a 3-bit fine-time measurement with a resolution of 1 ns within each 127.216~MHz system clock period. 
The wire data are packaged in the FPGA and distributed to both the DAQ and L1 trigger back-end systems through an SFP link with one GTX transceiver~\cite{gtx} and an AVAGO QSFP link with four GTX transceivers, respectively.

\subsection{Tracking in Level-1 trigger with CDC}

\begin{figure}[t]
\centerline{\includegraphics[width=1.5in]{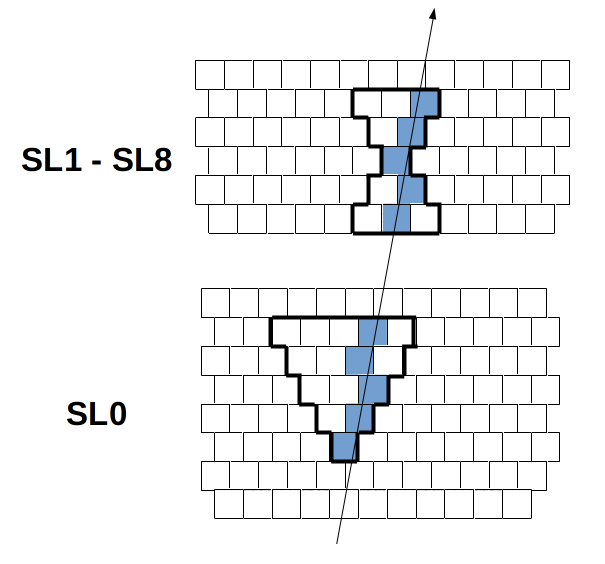}}
\caption{The shapes of the track segments for tracking with CDC in Level-1 trigger system.}
\label{fg:tsf}
\end{figure}

The design of the CDC L1 trigger system is described in detail in Ref.~\cite{cdctrg1,cdctrg2,cdctrg3}. 
The 292 CDC FEE boards transmit wire hit and TDC timing information to the trigger back-end system. Most of the trigger algorithms are implemented on Universal Trigger (UT) boards based on Xilinx Virtex-6 (UT3)~\cite{virtex6} and Virtex UltraScale (UT4)~\cite{virtexultrascale} FPGAs. The tracking chain of L1 trigger consists of the following modules:

\begin{enumerate}
\item Track Segment Finder (TSF): Nine TSF modules corresponding to the nine SLs of CDC receive data from each SL and reconstruct track segments from predefined hit patterns, as illustrated in Fig.~\ref{fg:tsf}. 
\item 2D Finder: Four 2D Finder modules reconstruct tracks in the transverse plane using track segments from the five axial SLs. A Hough-transform-based algorithm is used to determine the transverse momentum and incident angle of charged particles. Each 2D Finder covers one quarter of the CDC.
\item 3D Tracker: Using the 2D track candidates together with track segments from the four stereo SLs, the 3D Tracker measures the polar angle and the longitudinal offset relative to the Interaction Point (IP).
\item Neural 3D Tracker: In parallel with the fitter-based 3D Tracker, a Neural 3D Tracker employs pre-trained neural networks implemented in FPGA firmware to estimate the longitudinal track parameters~\cite{nn}.
\item Short Tracker: The Short Tracker reconstructs tracks that do not traverse the full CDC volume, such as low-momentum curling tracks or tracks toward the end-cap regions. It is based on pattern recognition with track segments reconstructed in the inner SLs.
\end{enumerate}
The reconstructed track segments and tracks from these modules are summarized by the Global Reconstruction Logic (GRL)~\cite{grl} and Global Decision Logic (GDL) into the final L1 trigger decision.

\subsection{Cross-talk noise within the Front-End Electronics device of CDC}

\begin{figure}[t]
  \centering
  \includegraphics[width=2.0in]{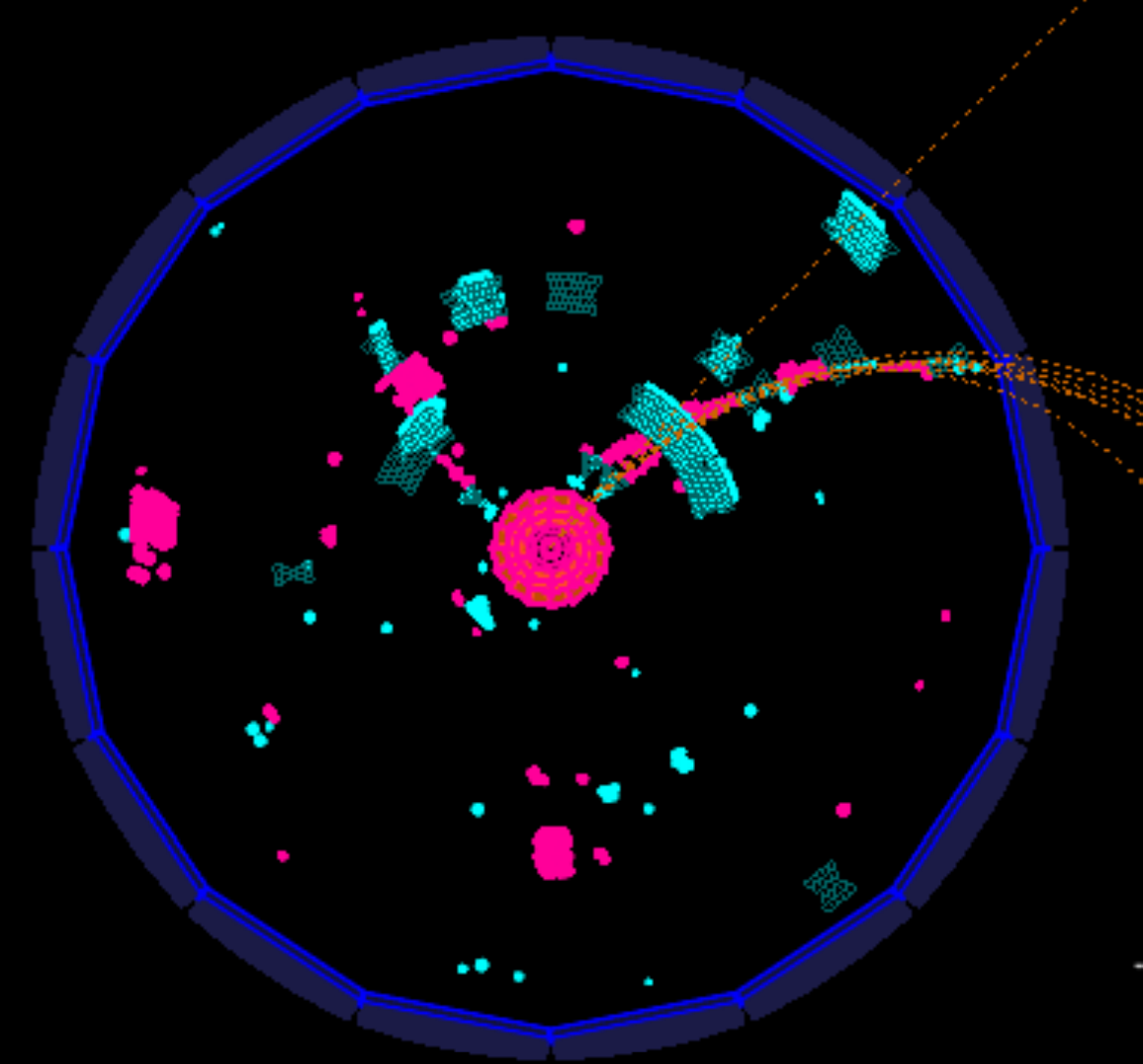}
  \caption{A Belle II event display demonstrating the cross-talk noise, where the event was taken in the Belle II beam collision runs in 2019. The cyan dots represent CDC wire hits in axial SLs. The magenta dots represent CDC wire hits in stereo SLs. The dotted curves in orange represent reconstructed L1 trigger tracks.}
  \label{fg:crosstalk}
\end{figure}

The cross-talk noise phenomenon in the CDC FEE has been a long-standing issue affecting Belle II operation. As shown in Fig.~\ref{fg:crosstalk}, which is an event display taken in the Belle II beam collision runs in 2019, multiple wire hits were observed in localized regions due to large energy deposits. A feature of these noise hits is that almost all eight wire channels within a single ASIC are fired simultaneously. 
The precise origin of the phenomenon has not yet been fully understood. The phenomenon is believed to originate within the CDC FEE and is also observed to depend on the Belle II electronics hut environment and beam background levels. One possible mechanism is related to the single-ended architecture of the ASD ASIC, where a large signal in one channel may induce correlated responses in neighboring channels through common electronic components.

The major impact of the cross-talk noise on the Belle II real-time data taking is the tracking performance in L1 trigger. Since the trigger tracking chain relies on reconstructed track segments rather than individual wire hits, it is particularly sensitive to cross-talk-induced fake segments.  
For the 2D Finder and the Short Tracker, fake tracks would be easily reconstructed due to noise-induced segments. 
For the 3D trackers, the noise affects not only the fake-track rate but also the resolution of the measured longitudinal track parameters. 
A study in Ref.~\cite{hough} indicates that the L1 trigger rate caused by fake tracks may reach $\mathcal{O}(100~\mathrm{kHz})$ under future beam-background conditions, presenting a significant challenge to trigger and DAQ operations at higher luminosities.

\subsection{Machine Learning in Front-End Electronics for noise suppression}

Various approaches have been investigated to mitigate the impact of cross-talk noise in the L1 tracking trigger system. 
For the Hough-transform-based 2D Finder, methods such as merging neighboring peaks in the conformal plane~\cite{grl} and extending the Hough transformation to wire-level information~\cite{hough} have been applied. 
In addition, graph-neural-network (GNN) based approaches have been explored for noise filtering over CDC~\cite{gnn1}. These methods are intended for deployment in the trigger back-end system, where devices equipped with larger FPGA chips are available.

In contrast, the present work targets suppression at the origin of the problem, namely the CDC FEE. 
Although the FPGA resources available in the Virtex-5 FPGA chip of FEE are significantly more limited than those of the trigger back-end system, suppression at the FEE level prevents noise from propagating downstream through the tracking chain. Furthermore, the FEE possesses unique information that is unavailable elsewhere in the trigger back-end system. The complete ADC waveform is accessible only within the CDC FEE, whereas the trigger back-end normally receives only reduced information such as wire-hit flags and TDC timing information.

These considerations motivate us to develop a ML-based waveform processing implementation in FEE. Using recorded waveform data of CDC, data-driven training is performed to develop ML models, and the models are converted into Register-Transfer-Level (RTL) modules for deployment in CDC FEE's Virtex-5 FPGA. The digitized ADC waveform will be processed in the FEE's FPGA in real-time to classify the hit as signal-like or noise-like, and the classification result is transmitted to the trigger back-end system for use in the trigger tracking algorithms. 

Data processing in the trigger system is fully pipelined without any deadtime, and has to satisfy the overall latency requirement of 4.4~$\mu$s. This new approach is expected to introduce a latency of less than 100~ns. Besides, the Virtex-5 FPGA of CDC FEE contains 97,280 Look-Up Tables (LUT), 97,280 Flip-Flops (FF), and 128 DSP slices, where the available resources are considerably more limited compared to modern FPGA families. These constraints impose stringent requirements on model complexity, latency, and FPGA resource consumption.
The ML models must be extremely compact and ultra-fast after RTL inference. We adopted a channel-by-channel implementation, where independent ML models process the waveform from each wire channel in parallel. This approach minimizes processing latency and the deadtime due to buffering data from each channel, while simultaneously requiring the deployed ML models to be much more compact. 

Under these constraints, several ML approaches were investigated, and a compact and ultra-fast Boosted-Decision-Tree-based solution was selected for deployment in the CDC FEE.

\section{Machine Learning Model development}
\label{sec:ml}

\subsection{ADC waveform data}

\begin{figure}[t]
  \centering
  \includegraphics[width=2.5in]{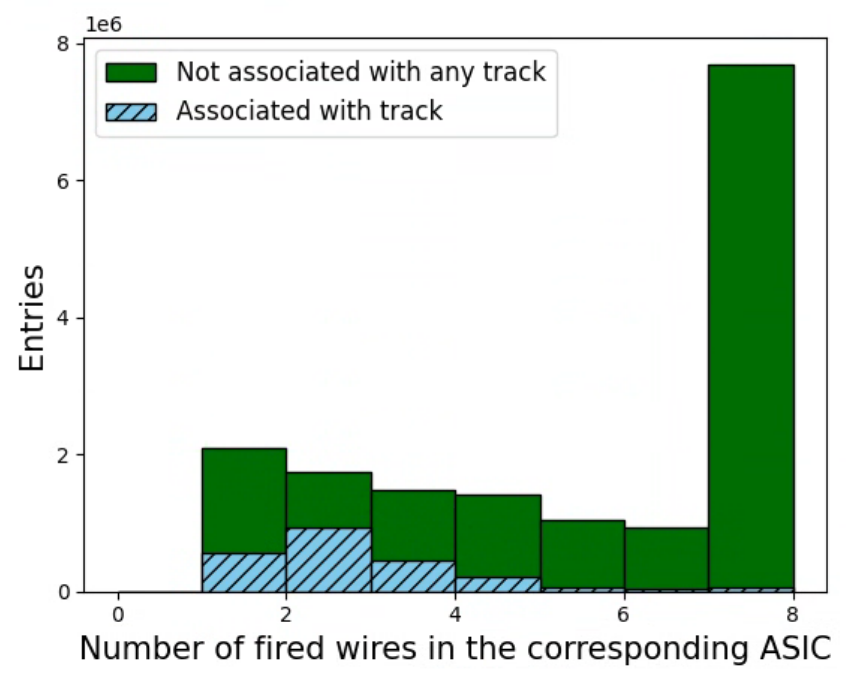}
  \caption{Number of fired wires in the corresponding ASIC.}
  \label{fg:nhits}
\end{figure}

Before developing the ML models, the CDC ADC waveform data were investigated to understand the characteristics of signal and noise-induced wire hits. Since the standard CDC readout does not record full ADC waveforms during beam collision data taking, a dedicated calibration run with waveform readout enabled was performed in 2024. The waveform data were sampled at 31.4~MSPS. With the aid of the Belle II offline reconstruction software~\cite{off1,off2,off3}, wire hits were categorized according to whether they were associated with reconstructed tracks. About 12.4\% of the wire hits are associated with reconstructed tracks. Figure~\ref{fg:nhits} shows the number of fired wires in the corresponding ASIC. The case in which all eight wire channels are fired within an ASIC occupies the majority of those not associated with any reconstructed tracks, which is consistent with our observation on the cross-talk noise. 

\subsection{Model building}
\label{sec:ml_model}

\begin{figure}[t]
  \centering
  \subfigure[Wire hits associated with reconstructed tracks]{
    \includegraphics[width=0.25\textwidth]{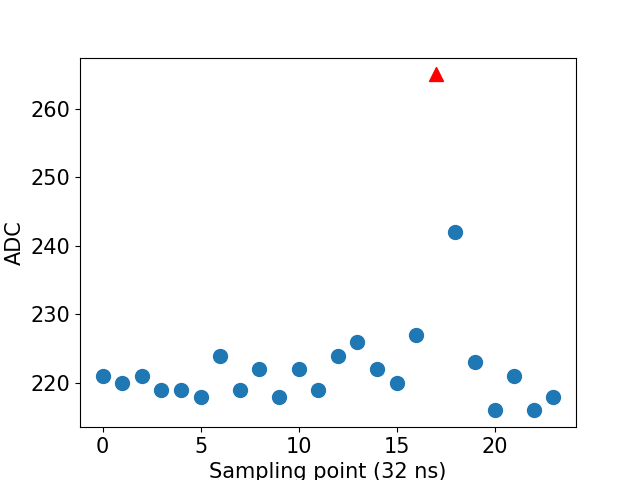}
    \includegraphics[width=0.25\textwidth]{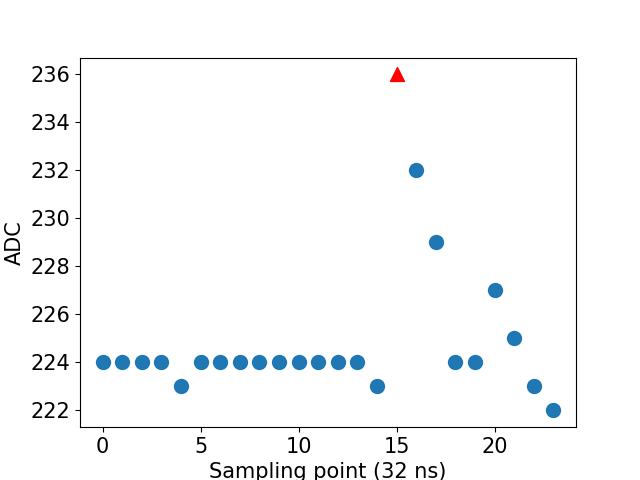}
    }
  \subfigure[Wire hits not associated with any reconstructed tracks]{
    \includegraphics[width=0.25\textwidth]{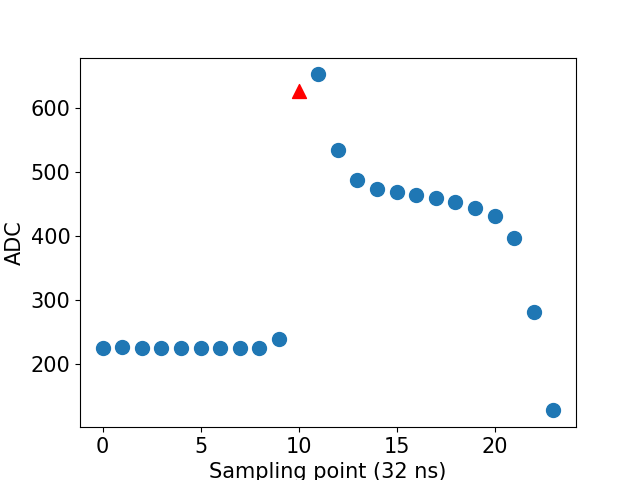}
    \includegraphics[width=0.25\textwidth]{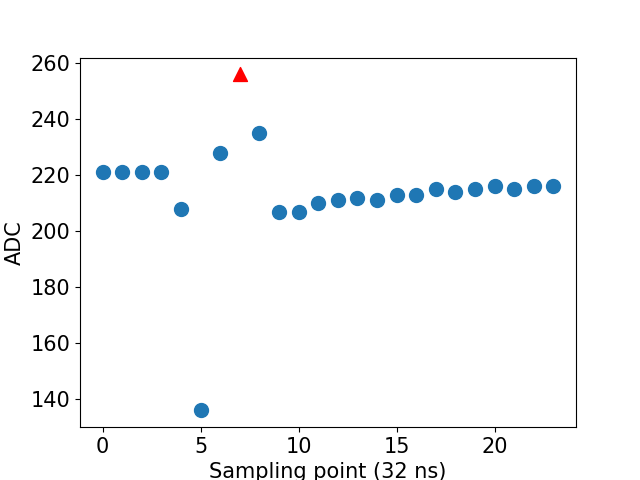}
    }
  \caption{Examples of ADC waveform of fired wire channels in CDC. The triangle points correspond to the TDC timing positions.}
  \label{fg:wave_ex}
\end{figure}

Figure~\ref{fg:wave_ex} shows examples of the ADC waveforms. For model training, the signal category consists of wire hits associated with reconstructed tracks and with fewer than four fired wires in the corresponding ASIC. The background category consists of wire hits not associated with any reconstructed track and with more than six fired wires in the corresponding ASIC. These requirements were introduced to obtain relatively pure signal and cross-talk-noise samples for training. As an initial feasibility study, both Neural Network (NN) and Boosted Decision Tree (BDT) classifiers with simple architectures were trained using the ADC waveform sampling points as input features. The BDT classifiers consistently achieved better performance, with the area under the ROC curve (AUC) being about 5--10\% higher than that of the NN models. Combined with their lower FPGA resource utilization and shorter latency for RTL inference, we adopted BDTs throughout this work.

\begin{figure}[t]
  \centering
  \includegraphics[width=2.0in]{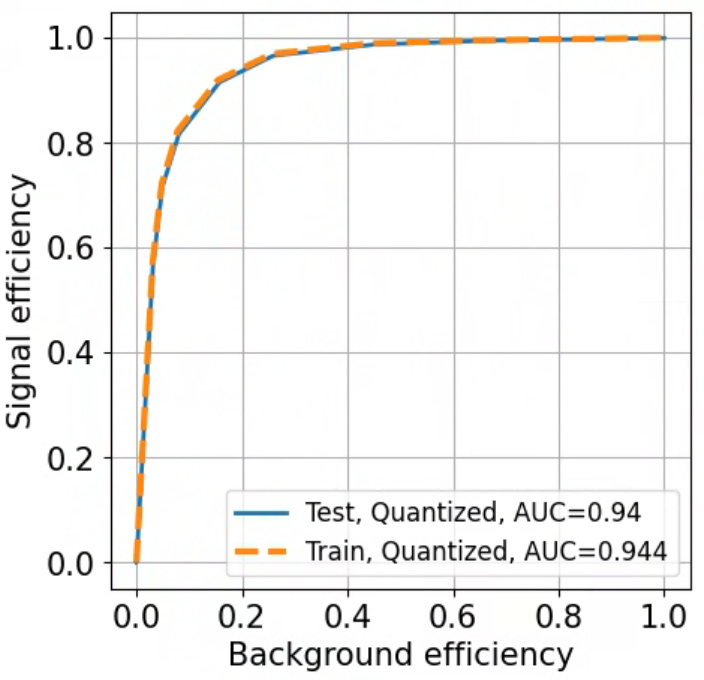}
  \caption{The receiver operating characteristic curve of one of the 96 quantized models.}
  \label{fg:roc}
\end{figure}

Five ADC sampling points centered on the TDC timing position were used as input features. Approximately 10,000 samples were selected for each of the signal and background categories. Only samples containing a single TDC hit within the selected timing window were used for training to avoid ambiguities arising from overlapping pulse structures. To further reduce the model complexity and FPGA resource usage, only the eight most significant bits (MSBs) of each ADC sampling point's value were retained. The BDT models were built using the \texttt{GradientBoostingClassifier} in the \texttt{scikit-learn} package and subsequently quantized using the Conifer package~\cite{conifer}. Four decision trees with a maximum depth of four and a learning rate of 1.0 were adopted as the model configuration. To account for geometric and electronic variations among channels within a FEE, independent models were trained for each of the 48 wire channels. In addition, since SL0 forms a standalone small cell chamber with a different cell geometry from the others, separate sets of models were trained for SL0 and for SL1--SL8. Hence, a total of 96 BDT models were developed. Figure~\ref{fg:roc} shows the receiver operating characteristic (ROC) curve for one channel. Only modest channel-to-channel variations are observed, with AUC values of 0.957--0.971 for SL0 and 0.934--0.954 for SL1--SL8. The differences in AUC between the \texttt{scikit-learn} model and the quantized model are below 1\% for all the 96 models, indicating negligible degradation in performance due to quantization.

\subsection{RTL Inference for the BDT Models and FPGA Deployment}

The Conifer package~\cite{conifer} supports converting BDT models into Xilinx High-Level-Synthesis (HLS) or VHDL codes. However, the Xilinx HLS software does not support FPGA families earlier than the 7 series, including the Virtex-5 FPGA used in the CDC FEE. A custom VHDL implementation of the BDT inference was developed to minimize both processing latency and FPGA resource utilization. The architecture is based on the fixed structure (four decision trees with a maximum depth of four) of the BDT models. In the first pipeline stage, all $4\times(2^{0}+2^{1}+2^{2}+2^{3}+2^{4})$ decision-node conditions from all the trees are evaluated in parallel and registered, requiring one clock-cycle. In the second stage, all the $4\times 2^{4}$ output conditions corresponding to each leaf node are evaluated by combining the corresponding decision-node outputs using AND gates and are subsequently registered, requiring an additional clock-cycle. Finally, the four classifier outputs from the four trees are summed combinationally to produce the final classifier without additional latency. The overall latency is 15.7~ns (two clock-cycles based on a 127.217 MHz system clock).
Based on the comparison in Table~\ref{tb:fpga}, our custom design achieves both the lowest latency and resource consumption. When all the 48 BDT modules are instantiated in the CDC FEE firmware, 4.7\% of the available FF and 4.4\% of the available LUT in the Virtex-5 FPGA were occupied. In addition, the measured increase in power consumption of the CDC FEE device is approximately 2\%.

\begin{table}
\caption{Comparison between the RTL inference designs. The Conifer HLS result was obtained based on Xilinx Kintex-7 FPGA since the Xilinx HLS software does not support Virtex-5 FPGA.}
\centering
\setlength{\tabcolsep}{3pt}
\begin{tabular}{c|c|c|c}
Inference  & \multirow{2}{*}{FF} & \multirow{2}{*}{LUT} & Processing Latency \\
Method &  &  & (clock-cycle) \\
\hline
Conifer HLS & 104 & 1876 & 7 \\
Conifer VHDL & 274 & 293 & 9 \\
Custom VHDL & 96 & 90 & 2 \\
\end{tabular}
\label{tb:fpga}
\end{table}

\section{Performance validation}
\label{sec:perf}

\subsection{Wire-Level Performance Based on Offline Waveform Data}

We first evaluated the performance of the BDT models at the individual wire-hit level. To define common selection criteria for all 96 quantized BDT models, the classifier thresholds were determined from their ROC curves. The threshold corresponding to a signal efficiency of approximately 99\% was defined as \textit{cut level 1}. Because the BDT classifier outputs are quantized, cut levels 2--4 were subsequently defined by consecutive increments of the classifier threshold.
Unlike the training samples described in Section~\ref{sec:ml_model}, we defined signal and background wire hits according to their association with reconstructed tracks to evaluate their efficiency losses due to the BDT selections. In addition, a dedicated category for cross-talk noise was further defined as background wire hits with more than six fired wires in the corresponding ASIC.
Although only single-TDC samples were used during training, the performance was evaluated using all waveform samples.
For waveform samples containing multiple TDC hits within the readout window, a separate set of five sampling points was constructed for each TDC timing position and processed independently by the BDT models. A wire hit was considered to pass the selection if at least one of the corresponding classifier outputs exceeded the selection threshold. The wire-level performance is summarized in Table~\ref{tb:wire_result}. With a drop in signal wire efficiency within 2\%, the cross-talk noise could be reduced by approximately 50\%.

\begin{table}
\caption{Wire-level performance obtained from offline processing of waveform data using the quantized BDT models, where the rejection rate is defined as 1.0 -- efficiency.}
\centering
\setlength{\tabcolsep}{3pt}

\begin{tabular}{c|ccc|ccc}
& \multicolumn{3}{|c|}{SL0} & \multicolumn{3}{|c}{SL1--SL8} \\
Cut & Signal & Background & Cross-talk& Signal & Background & Cross-talk\\
Level & efficiency & rejection rate & rejection rate & efficiency & rejection rate & rejection rate \\
\hline
1 & 0.993 & 0.271 & 0.441 & 0.995 & 0.149 & 0.233 \\
2 & 0.981 & 0.416 & 0.599 & 0.987 & 0.325 & 0.471 \\
3 & 0.959 & 0.543 & 0.728 & 0.973 & 0.513 & 0.698 \\
4 & 0.913 & 0.675 & 0.823 & 0.951 & 0.633 & 0.832 \\
\end{tabular}

\label{tb:wire_result}
\end{table}

\subsection{Real-Time Experimental Validation with FPGA Deployment}

The design was validated by deployment in the CDC FEE FPGA firmware in dedicated Belle II calibration runs taken in March 2026. 
Since such calibration tests with major modifications to the FEE firmware and L1 trigger configuration cannot be performed during regular beam-collision operation, they were carried out during special operation periods with only the electron beam at a beam current of 1~A, where CDC wire hits and charged tracks were still available, providing realistic experimental conditions to validate the noise-reduction design. 
Due to the limited operation time for these runs, three operating conditions were tested, including no ADC waveform requirement, BDT cut level~1, and BDT cut level~2, with a total of 341k, 222k, and 369k events collected, respectively.

In the CDC FEE, the TDC hit and timing information are available 251.5~ns earlier than the corresponding ADC waveform point because of the internal latency of the ADC and ASD ASICs, which is a key consideration when integrating the FEE firmware design within the L1 trigger system. Directly delaying the TDC information until the ADC classification becomes available would introduce an undesirable increase in the overall processing latency of the L1 trigger. 
To avoid this, the TDC information is transmitted immediately to the downstream Track Segment Finder boards to initiate segment reconstruction. Meanwhile, the corresponding ADC waveform is processed by the BDT modules. The classification output is subsequently transmitted to the Track Segment Finder and is matched to the associated wire hit during segment reconstruction. The segment finder then determines whether the segment should be retained or discarded according to the matching result. This processing scheme minimizes the additional latency from buffering TDC information.

\subsubsection{Level-1 Track Trigger Rate}

With the suppression of wire hits by the BDT selection, a corresponding reduction is expected in the numbers of reconstructed track segments as well as the trigger tracks. Therefore, the rates of the L1 track triggers provide a direct measure of the noise-suppression performance. 
Table~\ref{tb:trg_rate} summarizes the trigger rates measured during the calibration runs. The values were obtained from monitoring data archived by the Belle II slow control system based on EPICS~\cite{slc,trgslc} and are reported as the mean values with their standard deviations. Two types of full track triggers are shown: the 2D Finder using segments from the five axial SLs (SL0, 2, 4, 6, and 8), and the Neural 3D Tracker using segments from all nine SLs.
The track trigger rates are reduced by approximately 48\% and 72\% for the 2D Finder at cut levels 1 and 2, respectively. A smaller but still significant reduction is also observed for the Neural 3D Tracker.

\begin{table}
\caption{Level-1 trigger rates of the track triggers under the test conditions of the calibration runs.}
\centering
\setlength{\tabcolsep}{3pt}

\begin{tabular}{c|cc}
Cut& 2D Finder& Neural 3D Tracker \\
Level & (kHz) & (kHz)\\
\hline
None & $82.6\pm 1.8$ & $5.3\pm 0.1$\\
1 & $43.1\pm 0.9$ & $4.1\pm 0.1$\\
2 & $23.3\pm 1.1$ & $3.2\pm 0.1$\\
 
\end{tabular}

\label{tb:trg_rate}
\end{table}

\subsubsection{Track Segment Rate}

\begin{figure}[t]
  \centering
  \includegraphics[width=0.35\textwidth]{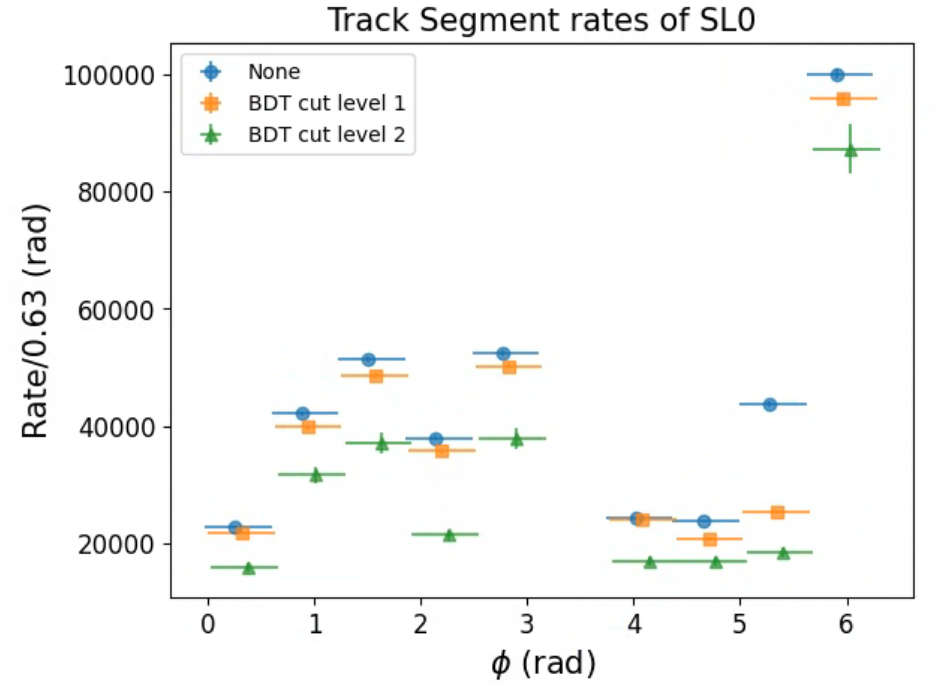}
  \includegraphics[width=0.35\textwidth]{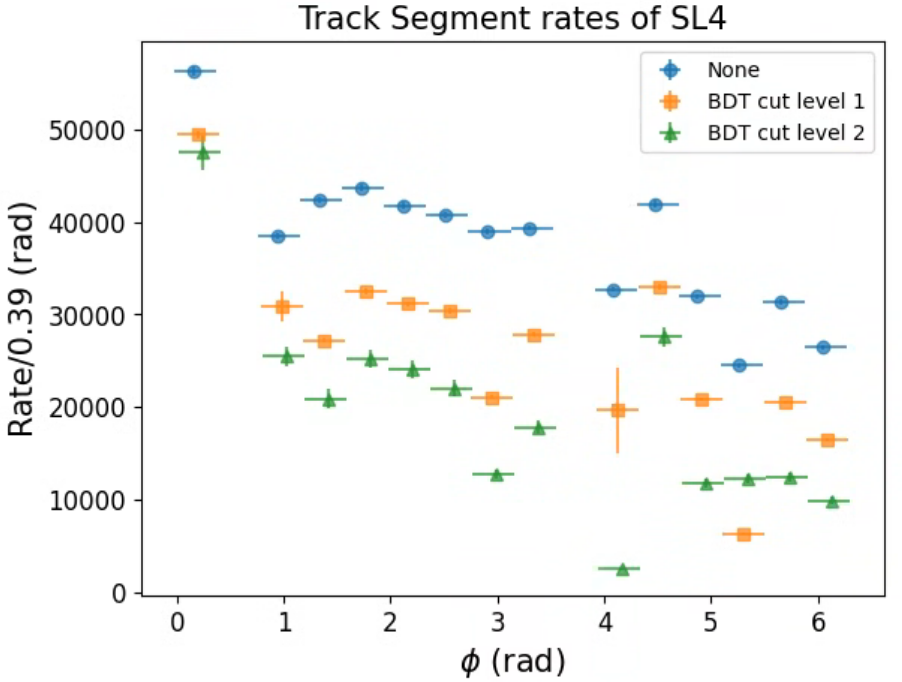}
  \includegraphics[width=0.35\textwidth]{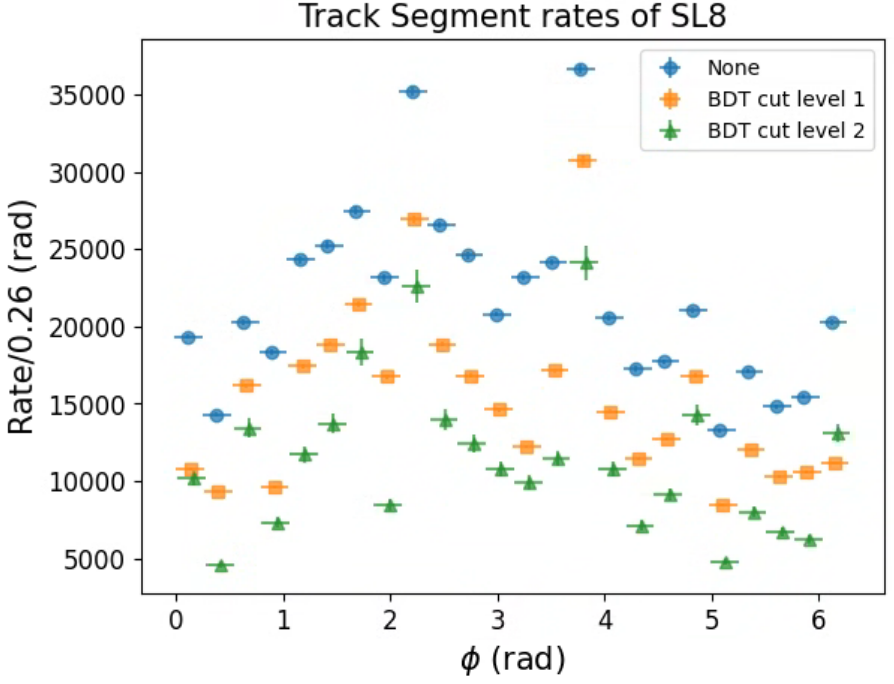}
  \caption{Track segment rates as a function of the azimuthal angle ($\phi$) for SL0, SL4, and SL8. The missing bin corresponds to detector regions masked during operation because of hot or dead channels.}
  \label{fg:ts_rate}
\end{figure}

Although the Belle II slow control system does not monitor individual CDC wire-hit rates, it records the rates of reconstructed track segments, which serve as another direct indicator of the noise-suppression performance. Table~\ref{tb:ts_rate} summarizes the fraction of track segment rates remaining after applying the BDT selection relative to the default condition of each SL. Figure~\ref{fg:ts_rate} shows the segment rates as a function of the azimuthal angle ($\phi$) for three representative super-layers, SL0, SL4, and SL8, where each bin corresponds to the coverage of 16 track segments (two FEEs). A substantial reduction in segment rates is observed for all super-layers, with the effect being particularly significant in the outer SLs, where an approximately 50\% reduction is observed at cut level~2. 

\begin{table}
\caption{Fraction of track segment rates remaining after the BDT selection relative to the default condition of each SL. The statistical uncertainties are below 4\% for all measurements.}
\centering
\setlength{\tabcolsep}{3pt}
\begin{tabular}{c|ccccccccc}
Cut & SL & SL  & SL  & SL  & SL  & SL  & SL  & SL  & SL  \\
Level & 0 & 1 & 2 & 3 & 4 & 5 & 6 & 7 & 8\\
\hline
1 & 0.962 & 0.935	& 0.775	& 0.806	& 0.749	& 0.733	& 0.749	& 0.694	& 0.702 \\
2 & 0.803 & 0.819	& 0.482	& 0.617	& 0.543	& 0.529	& 0.520	& 0.460	& 0.524 \\ 
\end{tabular} 

\label{tb:ts_rate}
\end{table}

\subsubsection{Track Trigger Acceptance}

To examine whether the signal wire hits which come from charged tracks are correctly retained after applying the BDT selection, the trigger acceptance of the track triggers was estimated using the recorded events. Events triggered by the calorimeter trigger were selected, and further required to contain at least one reconstructed full track from the IP (with $\Delta r <$ 10~cm and $\Delta z <$ 10~cm). Another category of events without any reconstructed tracks was also selected and considered as background events. 
Table~\ref{tb:trg_eff} summarizes the triggered fractions by the 2D Finder and Neural 3D Tracker under the different BDT selections. For events containing reconstructed IP tracks, only a modest reduction in trigger acceptance is observed after applying the BDT selection. In contrast, the acceptance for events without reconstructed tracks is reduced by approximately 40\%.

\begin{table}
\caption{Track-trigger acceptance under the test conditions of the calibration runs.}
\centering
\setlength{\tabcolsep}{3pt}
\begin{tabular}{c|cc|cc}
& \multicolumn{2}{|c}{Events with IP tracks} & \multicolumn{2}{|c}{Events without tracks} \\
Cut & 2D & Neural 3D & 2D & Neural 3D \\
Level & Finder & tracker & Finder & tracker \\
\hline
None & 0.944$\pm$0.032 & 0.790$\pm$0.032 & 0.277$\pm$0.001 & 0.044$\pm$0.001\\
1 & 0.935$\pm$0.046 & 0.744$\pm$0.039 & 0.182$\pm$0.001 & 0.028$\pm$0.001\\
2 & 0.896$\pm$0.032 & 0.702$\pm$0.027 & 0.173$\pm$0.001 & 0.028$\pm$0.001\\ 
\end{tabular} 

\label{tb:trg_eff}
\end{table}

\section{Prospect}
\label{sec:prospect}

\subsection{Future Extensions for CDC Cross-Talk Suppression}

In this study, we implemented independent waveform processing for each wire channel, and the BDT models demonstrated robust performance together with low latency and small FPGA resource consumption of RTL implementation. Future upgrades of the CDC FEE may enable more sophisticated ML architectures by considering correlations among neighboring channels. The coverage of wire channels within a FEE can be naturally represented as a two-dimensional spatial pattern, and the ADC sampling points provide an additional dimension. Such multi-dimensional information could be processed using convolutional neural networks (CNNs) or graph neural networks (GNNs) to further improve the discrimination between charged track signals and cross-talk noise. The FPGA chips with greater available resources are expected to enhance the feasibility of these approaches.

\subsection{Machine Learning in Detector Front-End Electronics}

This study demonstrates an implementation of ML in detector FEE under the stringent constraints of latency, power consumption, and FPGA resource utilization, and also highlights the potential for broader applications to similar use cases. Recent developments in high-energy physics instrumentation have increasingly considered ML techniques for processing detector information at the front-end level before data reduction takes place~\cite{ml}.
In many detector systems, waveform information is available only within the FEEs, whereas the trigger and DAQ back-end systems receive only extracted features, such as pulse height, timing, and other reduced quantities. 
Traditional waveform-processing techniques for feature extraction, such as constant fraction discrimination (CFD) and pulse shape fitting, can be sensitive to conditions like pile-up, overlapping pulses, and noise. 
ML-based methods provide an alternative and general approach to compact preprocessing in detector FEEs. With data-driven training, they are expected to suppress noise, separate overlapping pulses, and improve the extraction of timing, amplitude, and other waveform features, such that the performance of feature extraction on timing, pulse height, charge, and pulse-shape extraction can be further improved, and subsequently the performance of trigger algorithms in the back-end trigger system can be improved as well. 

The continued increase in FPGA capabilities, together with dedicated low-latency ML inference tools, is expected to expand the applicability of ML in detector front-end electronics. Such techniques may become increasingly important for future high-luminosity and high-rate experiments, where intelligent feature extraction and data reduction must be performed close to the detector before bandwidth-limited data transmission.

\section{Conclusion}
\label{sec:conclusion}
 
The Belle II Central Drift Chamber (CDC) has been affected by a cross-talk noise phenomenon in its front-end electronics (FEE), which degrades the performance of the Level-1 (L1) trigger tracking system. To address this issue, we developed and deployed an ML-based waveform processor in the CDC FEE for real-time cross-talk suppression. The design is based on Boosted Decision Trees (BDTs) with channel-specific RTL implementations and satisfies the stringent requirements on processing latency and FPGA resource utilization. 

Based on offline studies using recorded waveform data, the cross-talk noise can be reduced by approximately a factor of two while maintaining a signal-wire-hit efficiency above 98\%. The FEE FPGA firmware with the BDT modules deployed was further validated during dedicated Belle II calibration runs. Significant reductions in track-segment and track-trigger rates of up to approximately 50\% were observed, while the acceptance for events containing reconstructed charged tracks was maintained with a loss of less than 10\% loss.

This work demonstrates the technical feasibility of the direct deployment of compact, low-latency ML inference in detector FEEs. Beyond CDC cross-talk suppression, the approach provides a promising technical framework for ML preprocessing in future upgraded detector readout systems for high-energy physics experiments.

\section*{Acknowledgment}

The authors would like to thank the Belle II CDC and Level-1 Trigger groups for their continuous support of this work, especially in preparing and operating the dedicated calibration runs used for the experimental validation. 
The authors acknowledge the technical support provided by the Collider Electronics Forum of the Instrumentation Technology Development Center at KEK, together with its support for the underlying experimental infrastructure.
This work was supported by JSPS KAKENHI Grant Number JP25K01011.

\end{document}